\documentclass[fleqn,twoside]{article}
\usepackage[headings]{espcrc2}

\readRCS
$Id: espcrc2.tex,v 1.2 2004/02/24 11:22:11 spepping Exp $
\usepackage{graphicx}
\usepackage[figuresright]{rotating}
\usepackage{amssymb}
     \usepackage{bm}% bold math
     \usepackage{pifont}
      \usepackage{color}
      
      \newcommand{\BluTn}[1]{\textcolor{blue}{#1}}
      \newcommand{\RedTn}[1]{\textcolor{red}{#1}}
\newcommand{\AmS}{{\protect\the\textfont2
  A\kern-.1667em\lower.5ex\hbox{M}\kern-.125emS}}

\title{{\hfill RUB-TPII-03/08}\\ [1cm]
       New vistas of the meson structure in QCD from low to high
       energies
       }
\author{N.~G.~Stefanis\address[RUB]{Institut f\"ur
        Theoretische Physik II, Ruhr-Universit\"at Bochum,
        44780 Bochum, Germany}%
        \thanks{Talk presented at \emph{International Workshop
        on $e^+e^-$ collisions from Phi to Psi}, Frascati, Italy,
        7-10 April 2008.}}

\runtitle{New vistas of the meson structure in QCD}
\runauthor{N.~G.~Stefanis}

\begin{document}

\begin{abstract}
This talk presents issues pertaining to the quark structure
of the pion within QCD, both from the theoretical and from
the experimental point of view.
We review and discuss the pion-photon transition form factor
and the pion's electromagnetic form factor vs. corresponding
experimental data from the CLEO Collaboration and the JLab.
We also examine the extent to which recent high-precision lattice
computations of the second moment of the pion's distribution
amplitude conform with theoretical models.
Finally, we include predictions for the azimuthal asymmetry of
the $\mu^+$ distribution in the polarized $\mu$-pair-induced DY
production employing various pion distribution amplitudes.

\vspace{1pc}
\end{abstract}

\maketitle
\section{Introduction}
\label{sec:intro}
\begin{figure}[t]%[!thb]
\centerline{\includegraphics[width=0.9%
\columnwidth]{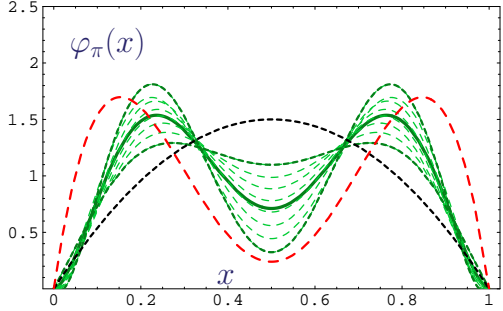}} % Alternative pionDAs.eps
\vspace*{-10mm}
\caption[*]{\footnotesize Pion's dilemma: to \textbf{B}
(Bactrian camel)---dashed line (CZ) \cite{CZ84} and BMS ``bunch''
\cite{BMS01}---or to \textbf{D} (Dromedary camel)---dotted line
(asymptotic DA)? All curves are shown at the normalization scale
$\mu^2=1$~GeV${}^2$.}
\label{fig:pion-dilemma}
\vspace*{-7mm}
\end{figure}

Understanding the quark structure of the lightest meson, the
pion,---the prototype for a meson bound state---is arguably one
of the most basic, albeit challenging, questions QCD is still
facing even after decades of intense investigations.
Integrating over transverse momenta up to some resolution scale
$\mu_{0}^{2}$, one gets from the pion's light-cone wave function the
pion's leading twist-2 distribution amplitude (DA),
$\varphi_{\pi}(x, \mu_{0}^{2})$, defined in terms of a nonlocal axial
current:
\begin{eqnarray}
  \langle 0 |\!\!\!\!\!\!\!\!\!\!\! && \bar{d}(z)
  \gamma^{\mu}\gamma_{5} {\cal C}(z,0) u(0)
  | \pi (P) \rangle |_{z^{2}=0}
 = \nonumber \\
 && \!\!\!\!\!\!\!\!\!
  if_{\pi}P^{\mu}\!\! \int_{0}^{1}\! dx\, {\rm e}^{ix(z\cdot P)}
\varphi_{\pi}^{\rm Tw-2}(x,\mu_{0}^{2}),
\label{eq:pi-DA}
\end{eqnarray}
%Eq (1)
where \hbox{$x$ ($\bar{x}\equiv 1-x$)} is the longitudinal momentum
fraction carried by the valence quark (antiquark) in the pion and the
path-ordered exponential
$
  {\cal C}(z,0)
=
  {\cal P} \exp \left[ -ig \int_{0}^{z} dy^{\mu} t^{a}A_{\mu}^{a}(y)
                \right]
$
ensures gauge invariance.
Appealing to its renormalization-group properties
\cite{ER80a,LB79}, we can expand
$\varphi_{\pi}^{\rm Tw-2}(x,\mu_{0}^{2})$ in terms of its one-loop
eigenfunctions, alias the Gegenbauer polynomials, to obtain
\begin{eqnarray}
   \varphi^{\rm Tw-2}(x; \mu_0^2)
\!\!\! \! && \!\!\!\!\!\! =
   \varphi^{\rm as}(x)
                      \left[1 + a_2(\mu_0^2)\ C^{3/2}_2(2x-1)
\right. \nonumber \\
&& \left.
 \! + \, a_4(\mu_0^2)\ C^{3/2}_4(2x-1)
       \right] + \ldots
\label{eq:pion-DA}
\end{eqnarray}
%Eq (2)
Here $\varphi^{\rm as}(x)=6x\bar{x}$ is the asymptotic
pion DA and by virtue of the leptonic decay
$\pi\to\mu^{+}\nu_{\mu}$ one has the normalization
$\int_{0}^{1} dx\ \varphi_{\pi}^{\rm Tw-2}(x,\mu_{0}^{2})=1$.
Relying only upon the first two Gegenbauer coefficients, Eq.\
(\ref{eq:pion-DA}) can yield distinct profiles, as shown in Fig.\
\ref{fig:pion-dilemma}.
We will see in the next section how two-photon processes can be used
to resolve pion's dilemma and reveal its parton substructure in
agreement with the experimental data and the latest lattice
calculations.
%%%%%%%%%%%%%%%%%%%%%%%%%%% F I G U R E  1 %%%%%%%%%%%%%%%%%%%%%%%%%%%%
%%%%%%%%%%%%%%%%%%%%%%%%%%%%%%%%%%%%%%%%%%%%%%%%%%%%%%%%%%%%%%%%%%%%%%%

\section{Pion DA, CLEO data, and lattice estimates}
\label{sec:DA-CLEO}

The entire nonperturbative content of the pion DA is encoded in the
expansion coefficients $a_n(\mu^2)$ which in turn can be derived from
the moments
\begin{equation}
  \langle \xi^{N} \rangle_{\pi}
\equiv
  \int_{0}^{1} dx (2x-1)^{N} \varphi_{\pi}(x)\, ,
\label{eq:moments}
\end{equation}
%Eq (3)
where $\xi\equiv 2x-1$, that decrease with increasing polynomial
order $N$ to 0:
$\langle \xi^{N} \rangle_{\pi}
\to
  [3/(N+1)(N+3)].
$
The evolution behavior of $a_n(\mu^2)$ is controlled by the
ERBL equation \cite{ER80a,LB79} (for a pedagogical
exposition, see \cite{Ste99}).
In our approach \cite{BMS01} we have determined $a_n$ at a
normalization scale $\mu_{0}^{2}=1.35$~GeV${^2}$ with the help of
QCD sum rules with nonlocal condensates \cite{MR89,BR91}.
There are alternative approaches, based, for instance, on local
QCD sum rules \cite{CZ84}, instantons \cite{PPRWG99,ADT00,PR01}, etc.

\begin{figure}[ht]%[!thb]
\centerline{\includegraphics[width=1.0%
\columnwidth]{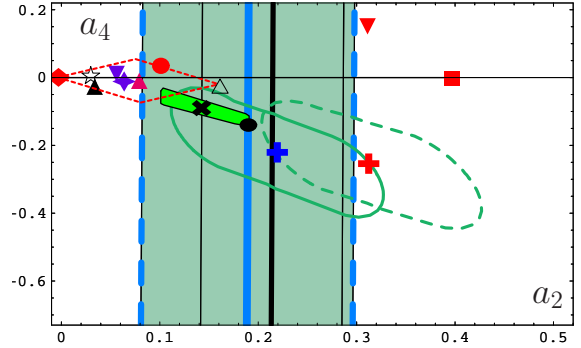}}
\vspace*{-10mm}
\caption[*]{\footnotesize Comparison at $\mu_{\rm CLEO}^2$ of various
pion DA models with the CLEO data in terms of the $1\sigma$ ellipse
(solid line) and recent lattice simulations, denoted by vertical
dashed lines \protect\cite{Lat06} and solid ones \cite{UKQCD-RBC07}.
The symbols mark the models discussed in Tables
\ref{tab:second_moment} and \ref{tab:fourth_moment}.
The slanted shaded rectangle represents the BMS DA ``bunch''
\protect\cite{BMS01}.
The dashed $1\sigma$ ellipse corresponds to the inclusion of the
twist-4 contribution to the pion DA via renormalons
\protect\cite{BMS05lat}.
The red bullet denotes the prediction derived from the holographic
model based on AdS/CFT duality (see last section).}
\label{fig:collage-CLEO-lattice}
\vspace*{-7mm}
\end{figure}

The predictive power of these theoretical approaches was challenged
by the high-precision data of the CLEO Collaboration on the pion-photon
transition \cite{CLEO98}, in which one photon has a large virtuality
$Q^2$, while the other is nearly on shell.
Kroll and Raulfs \cite{KR96} were the first to show that the popular
CZ pion DA was overshooting these data considerably.
Other analyses, having recourse to light-cone sum rules with a spectral
density obtained in the standard factorization scheme of perturbative
QCD in leading order (LO) \cite{Kho99} and next-to-leading-order (NLO)
\cite{SY99,BMS02,BMS03,BMS05lat}, followed, which established the
following facts (consult Fig.\ \ref{fig:collage-CLEO-lattice},
drawn at the main scale,
$\mu_{\rm CLEO}^2=\left(2.4~{\rm GeV}\right)^2$,
probed in the CLEO experiment):
(i) The CZ pion DA is outside the $4\sigma$ error ellipse of the CLEO
data.
(ii) The asymptotic pion DA is outside the $3\sigma$ ellipse, while
(iii) other proposed models close to that are at least $2\sigma$'s off.
(iv) The ``bunch'' of pion DAs derived from nonlocal QCD sum rules
mostly overlaps with the $1\sigma$ error ellipse, with the BMS model
DA being entirely inside.
%%%%%%%%%%%%%%%%%%%%%%%%%%% F I G U R E  2 %%%%%%%%%%%%%%%%%%%%%%%%%%%%
%%%%%%%%%%%%%%%%%%%%%%%%%%%%%%%%%%%%%%%%%%%%%%%%%%%%%%%%%%%%%%%%%%%%%%%
%%%%%%%%%%%%%%%%%%%%%%%%%%% F I G U R E  3 %%%%%%%%%%%%%%%%%%%%%%%%%%%%
\begin{figure}[h]%[!thb]
\centerline{\includegraphics[width=1.0%
\columnwidth]{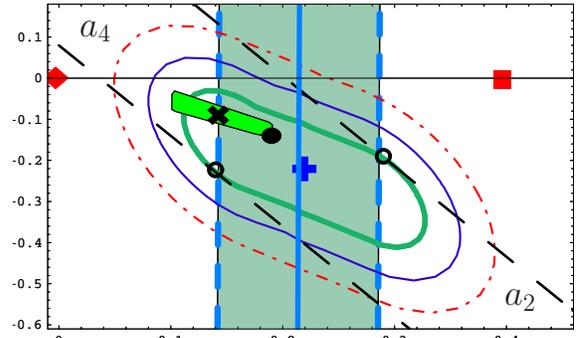}} \vspace*{-10mm}
\caption[*]{\footnotesize Range of values of $\langle
\xi^{4} \rangle_{\pi}(\mu_{\rm CLEO}^2)$ (slanted broken
lines) intersecting with the $1\sigma$ error ellipse (inner solid line)
of the CLEO data and the latest lattice results on $a_2$ of the
UKQCD-RBC Collaboration \cite{UKQCD-RBC07} (vertical shaded band).}
\label{fig:xi-4} \vspace*{-9mm}
\end{figure}
%%%%%%%%%%%%%%%%%%%%%%%%%%%%%%%%%%%%%%%%%%%%%%%%%%%%%%%%%%%%%%%%%%%%%%%

The most recent lattice calculations \cite{Lat06} (larger band
bounded by dashed lines) and \cite{UKQCD-RBC07} (narrower band within
solid lines) in Fig.\ \ref{fig:collage-CLEO-lattice} support and
enhance these findings as regards the range of values of the first
Gegenbauer coefficient $a_2$.
First, as one sees from this figure, they rather disfavor a
relatively large twist-4 contribution to the pion DA, estimated with
the help of renormalons \cite{BMS05lat} (dashed $1\sigma$ error
ellipse).
Second, the latest calculation \cite{UKQCD-RBC07}, with even smaller
uncertainties than \cite{Lat06}, indicates a trend further away
from the asymptotic DA, but still in compliance with the BMS results.

Tables \ref{tab:second_moment} and \ref{tab:fourth_moment} show the
values of the second and the fourth moment of various pion DAs at the
lattice reference scale $\mu_{\rm Lat}^{2}=4$~GeV${}^{2}$.

%%%%%%%%%%%%%%%%%%%%%%%%%%%%%%%%%%%%%%%%%%%%%%%%%%%%%%%%%%%%%%%%%%%%%%%
%%%%%%                          TABLE 1                          %%%%%%
%%%%%%%%%%%%%%%%%%%%%%%%%%%%%%%%%%%%%%%%%%%%%%%%%%%%%%%%%%%%%%%%%%%%%%%
\begin{table}[ht]
\caption{Predictions for $\langle \xi^{2} \rangle_{\pi}$ from
various models (methods) after NLO evolution from their proprietary
normalization point to the scale
$\mu_{\rm Lat}^{2}=4$~GeV${}^{2}$.
\label{tab:second_moment}}
\begin{tabular}{llll}\hline
Source       & Min   & Mid     & Max   \\ \hline
UKQCD/RBC \cite{UKQCD-RBC07} %
             & 0.252 & 0.278   & 0.304 \\
QCDSF/UKQCD \cite{Lat06} %
             & 0.230 & 0.269   & 0.308 \\
\cite{DelD05}  %DelDebbio
             & 0.232 & 0.285   & 0.338 \\ \hline
\cite{BMS02,BMS03} ($1\sigma$ CLEO)  \BluTn{\ding{58}} % BMS-LCSR
             & 0.240 & 0.280   & 0.320 \\
\cite{BMS01} ~{\ding{54}} % BMS-NLSR
             & 0.233 & 0.248   & 0.264 \\
\cite{SY99}  {\footnotesize\ding{108}} % Schmedding-Yakovlev
             & 0.233 & 0.269   & 0.305 \\ \hline
\cite{PR01}  \BluTn{\ding{70}} % Prasza{\l}owicz-Rostworowski
             & --- & 0.223     & ---   \\
\cite{ADT00} ~{\footnotesize\ding{115}} %ADT
             & --- & 0.212     & ---   \\
\cite{PPRWG99} ~{\ding{73}} % Petrov et al.
             & --- & 0.211     & ---   \\ \hline
\cite{Ag05a} {\footnotesize$\bigtriangleup$} % Agaev
             & --- & 0.259     & ---   \\
\cite{BZ05}  \RedTn{\footnotesize\ding{115}}% Ball-Zwicky
             & --- & 0.229     & ---   \\
\cite{BF89}  \RedTn{{\footnotesize\ding{116}}} % Braun-Filyanov
             & --- & 0.312     & ---   \\
\cite{CZ84}  ~~\RedTn{\footnotesize\ding{110}} % CZ
             & --- & 0.344     & ---   \\
Asy          \RedTn{\ding{117}}
             & --- & 0.199     & ---   \\ \hline
AdS/QCD      \RedTn{\ding{108}} & --- & 0.237 & ---    \\
\hline\end{tabular}
\end{table}
%%%%%%%%%%%%%%%%%%%%%%%%%%%%%%%%%%%%%%%%%%%%%%%%%%%%%%%%%%%%%%%%%%%%%%%

%%%%%%%%%%%%%%%%%%%%%%%%%%%%%%%%%%%%%%%%%%%%%%%%%%%%%%%%%%%%%%%%%%%%%%%
%%%%%%                          TABLE 2                          %%%%%%
%%%%%%%%%%%%%%%%%%%%%%%%%%%%%%%%%%%%%%%%%%%%%%%%%%%%%%%%%%%%%%%%%%%%%%%
\begin{table}[ht]
\caption{Predictions for $\langle \xi^{4} \rangle_{\pi}$ from
various models (methods) after NLO evolution from their proprietary
normalization point to the scale
$\mu_{\rm Lat}^{2}=4$~GeV${}^{2}$.
\label{tab:fourth_moment}}
\begin{tabular}{llll}\hline
      Source & Min     & Mid     & Max      \\ \hline
\cite{BMS02,BMS03} ($1\sigma$ CLEO)  %BMS-LCSR
             & 0.066   & 0.115   & 0.162    \\
\cite{BMS01} %BMS-NLSR
             & 0.105   & 0.108   & 0.113    \\
\cite{SY99}  %Schmedding-Yakovlev
             & 0.082   & 0.116   & 0.151    \\ \hline
\cite{PR01}  %Prasza{\l}owicz-Rostworowski
             & ---     & 0.10    & ---      \\
\cite{ADT00} %ADT
             & ---     & 0.091   & ---      \\
\cite{PPRWG99} %Petrov et al.
             & ---     & 0.094   & ---      \\ \hline
\cite{Ag05a} %Agaev
             & ---     & 0.121   & ---      \\
\cite{BZ05}  %Ball-Zwicky
             & ---     & 0.104   & ---      \\
\cite{BF89}  % Braun-Filyanov
             & ---     & 0.178   & ---      \\
\cite{CZ84}  % CZ
             & ---     & 0.181   & ---      \\
Asy
             & ---     & 0.085   & ---      \\ \hline
AdS/QCD      & --- & 0.114 & ---            \\
\hline\end{tabular}
\end{table}
%%%%%%%%%%%%%%%%%%%%%%%%%%%%%%%%%%%%%%%%%%%%%%%%%%%%%%%%%%%%%%%%%%%%%%%

In Fig.\ \ref{fig:xi-4}, we determine the range of values of the
fourth moment
$\langle \xi^{4} \rangle_{\pi}\left(a_{2},a_{4}|\mu^2\right)$
(denoted by the slanted broken lines)
that simultaneously fulfill the CLEO data and the lattice constraints
on $\langle \xi^{2} \rangle_{\pi}$ from
\cite{UKQCD-RBC07}.
For each error ellipse, there is some overlap with the vertical
$a_2$-band estimated in \cite{UKQCD-RBC07}.
For the maximum and minimum of
$\langle \xi^{4} \rangle_{\pi}$
with respect to the $1\sigma$ error ellipse, we find
\vspace{-0.9mm}
\begin{equation}
  0.095
\leq
  \langle \xi^4 \rangle_{\pi}\left(\mu_{\rm Lat}^{2}\right)
\leq
  0.134 \, ,
\label{eq:xi-4_min-max}
\end{equation}
%Eq (4)
corresponding to
$
  \left(a_{2}^{\rm min}
 =
  0.149, \;
  a_{4}^{\rm min}
=
  -0.241\right)
$ and
$
 \left(a_{2}^{\rm max}
 =
  0.307, \;
  a_{4}^{\rm max}
=
  -0.205\right).
$
[Recall that
$
 \langle \xi^{4} \rangle_{\pi}^{\rm asy}
=
 3/35.
$]
It happens that these values almost coincide with the intersection
points (open circles) of the $1\sigma$ error ellipse and the
boundaries of the vertical band of the $a_2$ lattice
constraints.

\begin{figure}[t]%[!thb]
\centerline{\includegraphics[width=1.0%
\columnwidth]{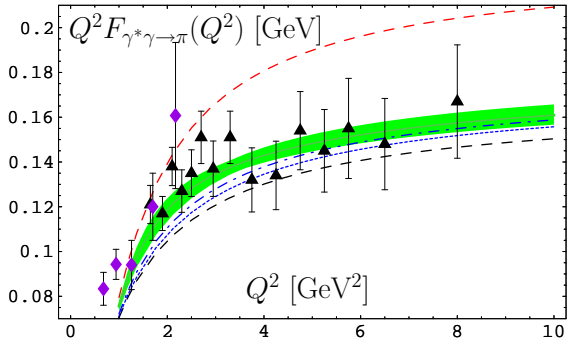}}
\vspace*{-10mm}
\caption[*]{\footnotesize Predictions for
   $Q^2F_{\gamma^*\gamma\to\pi}(Q^2)$ vs. $Q^2$ in comparison with
   available experimental data (see text).}
\label{fig:pi-gamma-ff}
\vspace*{-7mm}
\end{figure}

\begin{figure}[h]%[!thb]
\centerline{\includegraphics*[scale=0.2,angle=360,width=%
\columnwidth]{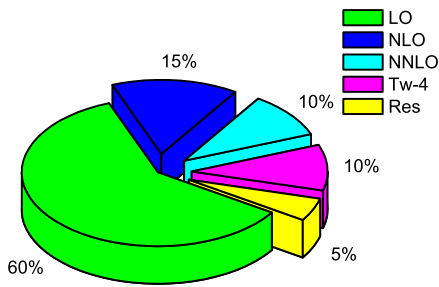}}
\vspace*{-15mm}
\caption[*]{\footnotesize Contributions to $Q^2F^{\gamma^{*}\gamma\pi}$
at the typical scale $\mu^{2}=5.76$~GeV${}^2$ of the CLEO data
\cite{CLEO98}.
The next-to-next-to-leading order (NNLO) estimate is based on
\cite{MMP02}, whereas the uncertainty owing to the (Res)onance model
used, was discussed in \cite{Kho99}.}
\label{fig:pi-gamma-pie}
\vspace*{-8mm}
\end{figure}

Figure \ref{fig:pi-gamma-ff} displays various theoretical predictions
for $Q^2F_{\gamma^*\gamma\to\pi}(Q^2)$ vs. $Q^2$ in comparison with the
CELLO (diamonds, \protect\cite{CELLO91}) and the CLEO (triangles,
\protect\cite{CLEO98}) experimental data, evaluated with the
twist-4 parameter value $\delta_{\rm Tw-4}^2=0.19$~GeV$^2$
\cite{BMS02,BMS03,BMS05lat}.
The other curves shown correspond to selected pion DAs:
the asymptotic DA $\varphi_{\rm as}$ (lower dashed line),
$\varphi_{\rm CZ}$ (upper dashed line) \cite{CZ84}, and
two instanton-based models, viz., \cite{PPRWG99} (dotted line)
and \cite{PR01} (dash-dotted line).
An important observation from this figure is that the shaded strip,
which corresponds to the BMS ``bunch'' \cite{BMS01}, becomes narrower
at lower scales around 1~GeV${}^2$.
The reason is that at such low scales, the form factor is dominated
by its twist-4 contribution, while the leading twist-2 part dies out.
Moreover, Fig.\ \ref{fig:pi-gamma-ff} makes it clear that the low-$Q^2$
CELLO data \cite{CELLO91} exclude $\varphi_{\pi}^{\rm asy}$ and clones,
while the high-$Q^2$ CLEO data \cite{CLEO98} rule out
$\varphi_{\pi}^{\rm CZ}$.
Figure \ref{fig:pi-gamma-pie} gives an illustration of the partial
contributions to $Q^2F^{\gamma^{*}\gamma\pi}$, originating from
different sources at $\mu_{\rm CLEO}^2$.
A comprehensive account of these effects can be found in
\cite{BMS05lat}.
%%%%%%%%%%%%%%%%%%%%%%%%%%% F I G U R E  4 %%%%%%%%%%%%%%%%%%%%%%%%%%%%
%%%%%%%%%%%%%%%%%%%%%%%%%%%%%%%%%%%%%%%%%%%%%%%%%%%%%%%%%%%%%%%%%%%%%%%
%%%%%%%%%%%%%%%%%%%%%%%%%%% F I G U R E  5 %%%%%%%%%%%%%%%%%%%%%%%%%%%%
%%%%%%%%%%%%%%%%%%%%%%%%%%%%%%%%%%%%%%%%%%%%%%%%%%%%%%%%%%%%%%%%%%%%%%%

\section{Electromagnetic pion form factor}
\label{sec:em-ff}

\begin{figure}[t]%[!thb]
\centerline{\includegraphics*[width=1.0\columnwidth]{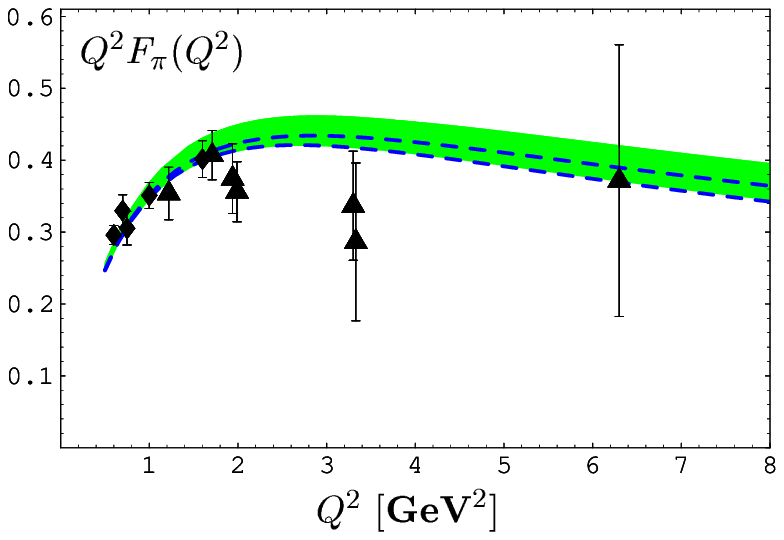}}
\vspace*{-10mm}
\caption[*]{\footnotesize Predictions for the scaled pion form
        factor calculated with the BMS ``bunch'' (shaded strip)
        \protect{\cite{BMS01}} in NLO QCD analytic perturbation
        theory \protect\cite{BPSS04}.
        The dashed lines inside the strip restrict the area of
        predictions accessible to the asymptotic pion DA.
        The experimental data are taken from \protect{\cite{JLab00}}
        (diamonds) and \cite{FFPI73} (triangles).}
\label{fig:pion-ff}
\vspace*{-0.7mm}
\end{figure}

The analysis of the pion's electromagnetic form factor involves on the
nonperturbative side the BMS ``bunch'' of pion DAs in comparison with
$\varphi_{\pi}^{\rm asy}$ and $\varphi_{\pi}^{\rm CZ}$.
On the perturbative side, a theoretical scheme is used, which consists
of expanding the form factor in terms of analytic images of the strong
running coupling and its powers up to the NLO.
The basis of this approach develops from \cite{SS97} and can account
for more than one hard scale by incorporating into the
``analytization'' procedure all terms that contribute to the spectral
density of the amplitude \cite{KS01}.
The bedrock of the approach was developed in
\cite{KS01,SSK99,Ste02,SSK00,BMS05,BMS06}
and its application to $F_{\pi}^{\rm NLO}(Q^2)$ was considered in
\cite{BPSS04,BKS05} (see also \cite{Bak08}).
The results are displayed in Fig.\ \ref{fig:pion-ff} in comparison
with experimental data.
Note that the quantity shown comprises a soft non-factorizing
contribution which dominates at currently accessible momentum
transfers \cite{BPSS04}.

The main characteristics of this approach are
(i) a strongly reduced sensitivity on the renormalization and the
factorization scale,
(ii) an undiminished quality of precision in adopting different
choices of renormalization schemes and scale settings, virtually
eliminating the dependence variations from scheme to scheme and scale
to scale \cite{BKS05}.
(iii) Another important finding is that, within such an analytic
approach, the form-factor predictions (shaded strip in
Fig.\ \ref{fig:pion-ff}) turn out to be very close to that computed
with $\varphi_{\pi}^{\rm asy}$, albeit the underlying pion DA profiles
are very different.
This proved that what really matters is the behavior of the pion DA
at the endpoints $x\to 0,1$.
%%%%%%%%%%%%%%%%%%%%%%%%%%% F I G U R E  6 %%%%%%%%%%%%%%%%%%%%%%%%%%%%
%%%%%%%%%%%%%%%%%%%%%%%%%%%%%%%%%%%%%%%%%%%%%%%%%%%%%%%%%%%%%%%%%%%%%%%

\section{Conclusions}
\label{sec:concl}
We have shown that the CLEO data on $F^{\gamma^*\gamma\pi}$ pose a veto
to a variety of proposed models for the pion DA
(see Fig.\ \ref{fig:collage-CLEO-lattice}) and favor an
endpoint-suppressed ''B''-shaped pion DA---a Bactrian ``camelino'', the
endpoint suppression being provided by the vacuum nonlocality
$\lambda_{q}^{2}=0.4$~GeV${}^2$.
The crucial `missing link' is the determination of
$\langle \xi^{4} \rangle_{\pi}$ on the lattice.
A value within the range $[0.095-0.134]$(2~{\rm GeV}), extracted
from the CLEO data in conjunction with the lattice constraints on
$\langle \xi^{2} \rangle_{\pi}$ from \cite{UKQCD-RBC07} (cf.\ Fig.\
\ref{fig:xi-4}), would validate the claim that the pion DA is BMS-like.

\begin{figure}[t]%[!thb]
\centerline{\includegraphics*[width=1.0\columnwidth]{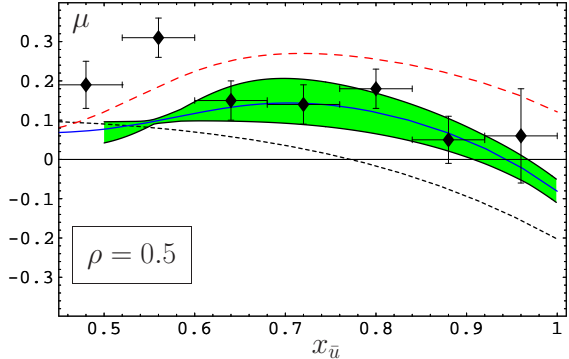}}
\vspace*{-10mm}
\caption[*]{\footnotesize Results for the angular distribution
   parameter $\mu$ as a function of
   $x_{\bar{u}}\equiv x_{\pi}$ for
   $\rho\equiv Q_{T}/Q=0.5$.
   The shaded strip contains the results for the BMS ``bunch'' of DAs
   \protect\cite{BMS01}.
   The solid line corresponds to the BMS model, the dotted solid
   denotes the result for
   $\varphi_{\pi}^{\rm asy}$, and the (red) dashed line is the
   prediction for the \emph{endpoint--dominated} CZ DA
   \protect\cite{CZ84}.
   One-loop evolution of the pion DAs to each measured $Q^2$ value is
   included (data taken from \protect\cite{Con89}).}
\label{fig:DY}
\vspace*{-9mm}
\end{figure}

The Drell-Yan process $\pi^{-}N\to \mu^{+}\mu^{-}X$ for
lepton-pair production with a large invariant mass $Q^2$ provides an
additional useful tool to probe and test different pion DAs in terms
of azimuthal asymmetries, as Fig.\ \ref{fig:DY} illustrates for the
case of the kinematic variable $\mu$.
%%%%%%%%%%%%%%%%%%%%%%%%%%% F I G U R E  7 %%%%%%%%%%%%%%%%%%%%%%%%%%%%
%%%%%%%%%%%%%%%%%%%%%%%%%%%%%%%%%%%%%%%%%%%%%%%%%%%%%%%%%%%%%%%%%%%%%%%
Overall, a rather good agreement of the BMS ``bunch'', derived from
nonlocal QCD sum rules \cite{BMS01}, with available data was found
\cite{BST07}, though the existing data cannot single out a particular
model.
In this respect, the planned COMPASS experiment may be of significant
relevance.

\section{Section added: Holographic dual of QCD}
\label{sec:AdS}

In this extra section, contained only in the arXiv version of this
paper, we provide some predictions from the AdS/QCD approach.

In the holographic model of Brodsky and de T\'{e}ramond
(see, e.g., \cite{BT08}), the pion DA has the form
\begin{equation}
  \varphi(x)_{\pi}^{\rm hol}
\sim
  \sqrt{x(1-x)}
\label{eq:BT-DA}
\end{equation}
%Eq (5)
with the normalization
\begin{equation}
  \int_{0}^{1}dx\ \varphi(x)_{\pi}^{\rm hol}
=
  B(3/2,3/2)
=
 \frac{\pi}{8}
\label{eq:BT-DA-norm}
\end{equation}
%Eq (6)
and moments given by
\begin{equation}
  \langle \xi^{2n} \rangle_{\pi}^{\rm hol}
=
  \frac{1}{4}\frac{B\left(3/2,(2n+1)/2\right)}{B(3/2,3/2)} \, ,
\label{eq:BT-DA-moments}
\end{equation}
%Eq (7)
where $B(x,y)$ is the Euler Beta function.
Taking the first derivative of $\varphi(x)_{\pi}^{\rm hol}$,
$\sim \frac{1}{\sqrt{x}}$, one realizes that this DA has its endpoints
strongly enhanced, even relative to the asymptotic DA, let alone to
the BMS DA (cf.\ Fig.\ \ref{fig:pion-dilemma}).
Note that also the twist-4 DA contribution, extracted from the
holographic model by Agaev and Gomshi Nobary \cite{Aga08}, shows
endpoint enhancement, although this effect is milder compared to the
twist-2 DA case.
It is interesting to recall at this point that Eq.\ (\ref{eq:BT-DA})
was considered long ago by Mikhailov and Radyushkin in an attempt to
reconstruct the pion DA from its first few moments within the context
of QCD sum rules with nonlocal condensates \cite{MR89}.

The values of the moments $\langle \xi^{2} \rangle_{\pi}$
and $\langle \xi^{4} \rangle_{\pi}$ of the holographic model have
been inserted in Tables \ref{tab:second_moment} and
\ref{tab:fourth_moment}, respectively, (last entry in each of these
tables) after NLO evolution from their initial values
$\langle \xi^{2} \rangle_{\pi}=1/4$ and
$\langle \xi^{4} \rangle_{\pi}=1/8$ to the lattice scale
$\mu_{\rm lat}^2$ (with associated Gegenbauer coefficients
$a_2^{\rm hol}(\mu_{\rm lat}^2)=0.107$
and
$a_4^{\rm hol}(\mu_{\rm lat}^2)=0.038$).

The holographic model is also included in Fig.\
\ref{fig:collage-CLEO-lattice} in comparison with other theoretical
models and confronted with the CLEO data \cite{CLEO98}, in parallel
with the two most recent and precise lattice calculations
\cite{Lat06,UKQCD-RBC07}.
As for the moments above, we have assumed that the holographic DA,
given by Eq.\ (\ref{eq:BT-DA}) is normalized at the scale
$\mu^2=1$~GeV${}^2$
with the Gegenbauer coefficients
$a_2^{\rm hol}(\mu^2=1~{\rm GeV}^2)= 7/48$
and
$a_4^{\rm hol}(\mu^2=1~{\rm GeV}^2)= 11/192$.
We then performed a two-loop evolution to the scale
$\mu_{\rm CLEO}^2$ and found
\begin{equation}
a_2^{\rm hol}(\mu_{\rm CLEO}^2)=0.101,~~~
a_4^{\rm hol}(\mu_{\rm CLEO}^2)=0.03 \ .
\label{eq:gegen-BT_CLEO}
\end{equation}
%Eq (8)
The position of the holographic model is marked in Fig.\
\ref{fig:collage-CLEO-lattice} by a red bullet and turns out to be
(cf.\ Fig.\ \ref{fig:xi-4})
just inside the border of the $2\sigma$ ellipse of the CLEO data,
being also inside the predictions' range of the QCDSF/UKQCD
Collaboration, but outside the limits on $a_2$ determined by the
UKQCD-RBC07 Collaboration.

Predictions for the spacelike (and timelike) pion form factor,
derived from AdS/QCD, can be found, for example, in
\cite{BT08,Aga08,GR07,KL07}; they are not addressed here.

\section*{Acknowledgments}
I wish to thank A.P.~Bakulev for collaboration and S.V.~Mikhailov
for collaboration and technical help in preparing this contribution.
I am indebted to the organizers of the workshop for financial support
and the Deutsche Forschungsgemeinschaft (DFG) for a travel grant.
The works presented here were supported in part by the DFG under Grant
436 RUS 113/881/0 and the Heisenberg-Landau Programme (Grant 2008).
Finally, I would like to thank S. Brodsky and G. de T\'{e}ramond for
attracting my attention to holographic QCD.

\end{document}